\documentstyle[aps,prl,multicol,psfig]{revtex}
\begin{document}
\draft
\title{Misleading signatures of quantum chaos}

\author{J. M. G. G\'omez, R. A. Molina, A. Rela\~no and  J. Retamosa}

\address{Departamento de F\'{\i}sica At\'omica, Molecular y Nuclear, 
         Universidad Complutense de Madrid,  
         E-28040 Madrid, Spain}

\maketitle

\begin{abstract}

The main signature of chaos in a quantum system is provided by
spectral statistical analysis of the nearest neighbor spacing
distribution and the spectral rigidity given by $\Delta_3(L)$.  It is shown
that some standard unfolding procedures, like local unfolding and Gaussian
broadening, lead to a spurious increase of the spectral rigidity
that spoils the $\Delta_3(L)$ relationship with the regular or chaotic 
motion of the system. This effect can also be misinterpreted as 
Berry's saturation.

\end{abstract}


\begin{multicols}{2}
\narrowtext
%
%

Quantum chaos has been an active research field since the link between
energy level fluctuations and the chaotic or integrable properties of
Hamiltonian systems was conjectured \cite{Berry:81,Bohigas:84},
providing one of the fundamental signatures of quantum chaos
\cite{Granada:84,Stockman:99} in atoms, molecules, nuclei, quantum
dots, etc.  The secular or smooth behavior of of the level density is
a characteristic of each quantum system, while the fluctuations
relative to this smooth behavior are related to the regular or chaotic
character of the motion in all quantum systems.  To achieve the
separation of the smooth and fluctuating parts, the energy spectrum is
scaled to a sequence with the same local mean spacing along the whole
spectrum. This scaling is called {\it unfolding}
\cite{Brody:81}. Although this can be a non-trivial task
\cite{Guhr:98}, the description of the unfolding details of
calculations is usually neglected in the literature.

In this Letter we show that, contrary to common assumptions, the
statistics that measure long range level correlations are strongly
dependent on the unfolding procedure utilized, and some standard
unfolding methods give very misleading results in regard to the
chaoticity of quantum systems. Long range level correlations are
usually measured by means of the Dyson and Mehta $\Delta_3$ statistic
\cite{Brody:81}. On the other hand, short range correlations,
caracterized by the the nearest-neighbor spacing distribution $P(s)$,
are not very sensitive to the unfolding method.

%
%

\begin{figure}[h]
\begin{center}
\leavevmode
\psfig{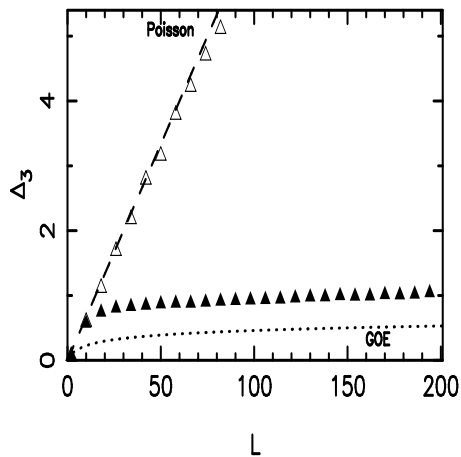}
\end{center}
\caption[]{Comparison of the $\Delta_3$ statistic for a
rectangular quantum billiard using two unfolding procedures. Open
triangles correspond to the smooth unfolding using the Weyl law, and
filled triangles to the local unfolding method.  The dashed line is
the Poisson limit and the dotted line is the GOE limit.}
\label{billiard}
\end{figure}

Let us consider a rectangular quantum billiard with a size ratio
$a/b=\pi$.  This is a well known example of a regular system. In
general, for regular systems level fluctuations behave like in a
sequence of uncorrelated energy levels, and the $\Delta_3(L)$
statistic increases linearly with $L$. However, it was shown by Berry
\cite{Berry:85} that the existence of periodic orbits in the phase
space of the analogous classical system leads to a saturation of
$\Delta_3(L)$ for $L$ larger than a certain value $L_s$, related to
the period of the shortest periodic orbit.  Fig. \ref{billiard} shows
the $\Delta_3$ behavior for a sequence of 8000 high energy levels of
the mentioned quantum billiard, calculated with two different
unfolding procedures. The mean level density for this system is given
by the Weyl law \cite{Haake:01}.  Using this density to perform the
unfolding, $\Delta_3$ follows the straight line of level spacings with
Poisson distribution, characteristic of regular systems. In this
example, Berry's saturation takes place at $L_s \simeq 750$, that is
outside the figure.  Let us suppose now that the law giving the mean
level density of the system were unknown. Then, a standard method to
obtain the local mean level density at energy $E$ is to calculate the
average density of a few levels around this energy. Using this method
one obtains a very different behavior, the spectral rigidity increases
strongly at $L\simeq 20$, and afterwards $\Delta_3$
is close to the Gaussian orthogonal ensemble (GOE)
line characteristic of chaotic systems. The latter behavior is not at
all related to the Berry saturation. It is a spurious effect due to
inappropriate unfolding of the level spectrum and it implies that
strong long-range correlations have been improperly introduced by the
procedure.

This first example illustrates the problem that can arise with some
reasonable unfolding methods currently used in quantum chaos
calculations \cite{French:71,Heyde:91,Otsuka:92,Meyer:97,Bruus:97}.
In order to understand its origin we shall analyze different
unfolding procedures.  The principal difficulty in the unfolding is
the correct characterization of the mean level density
$\bar{\rho}(E)$. Having this function, the unfolded adimensional
variables $\varepsilon_i$,
\begin{equation}
\varepsilon_i=\bar{N}(E_i),\;\;\;\;\bar{N}(E)=\int{dE\bar{\rho}(E)},
\label{densmedint}
\end{equation}
have  mean level density $\bar{\rho}(\varepsilon)=1$.  The unfolded
spacing sequence is then $\{s_i=\varepsilon_{i+1}-\varepsilon_i\}$,
and the nearest-neighbor spacing distribution $P(s)$ is well suited to
study the short-range spectral correlations \cite{Brody:81}.

The $\Delta_3$ statistic is used to investigate the long range correlations. 
It is defined for the interval $[a,a+L]$ in the cumulative level density as
\begin{equation}
\label{delta3}
\Delta_{3}(a,L) = \frac{1}{L}\,\,\, \min_{A,B} \int_{a}^{a+L}
\left[N(\varepsilon)-A \varepsilon-B\right]^2 d\varepsilon .
\end{equation}
The function $\Delta_{3}(L)$, averaged over intervals, measures the
deviations of the quasi-uniform spectrum from a true equidistant
spectrum.

For some systems a natural unfolding procedure exists, because
$\bar{\rho}(E)$ is known from an appropriate statistical theory or by a
well checked empirical ansatz. For example, $\bar{\rho}(E)$ is a
semicircle for large GOE matrices \cite{Brody:81}, it often has
Gaussian form for large nuclear shell-model matrices \cite{Brody:81},
and follows the Weyl law in quantum billiards \cite{Haake:01}.
However, in many systems where there is no natural choice for
$\bar{\rho}(E)$, it is usually estimated from a set of neighboring
levels. The simplest method, called {\it local unfolding}, has been
widely used \cite{French:71,Heyde:91,Otsuka:92,Meyer:97}. The mean
level density is assumed to be approximately linear in a window of
$v$ levels on each side of $E_i$, and is given by
\begin{equation}
\bar{\rho}_L (E_i)= \frac{2v}{E_{i-v}-E_{i+v}},
\label{rholoc}
\end{equation}
where $L$ stands for local unfolding.  More sophisticated is the {\it
Gaussian broadening} method \cite{Haake:01,Bruus:97}.  The level
density $\rho(E) = \sum_i \delta(E-E_i)$ is substituted by an average
level density
\begin{equation}
\bar{\rho}_G (E) = \frac{1}{\sigma\sqrt{2\pi}} \sum_i 
\exp\left\{-\frac{(E-E_i)^2}{2\sigma^2}\right\},
\end{equation}
where $G$ stands for Gaussian broadening. The sum runs over all the
energy levels, but only those satisfying $\left| E-E_i\right| \alt
\sigma$ do significantly contribute to $\bar{\rho}_G(E)$. Although
these two methods are different, both depend on a parameter $v$ or
$\sigma$ that measures, in a real or effective way, how many
neighboring levels are used to calculate the local mean density. 

%
%
\begin{figure}[h]
\begin{center}
\leavevmode
 \psfig{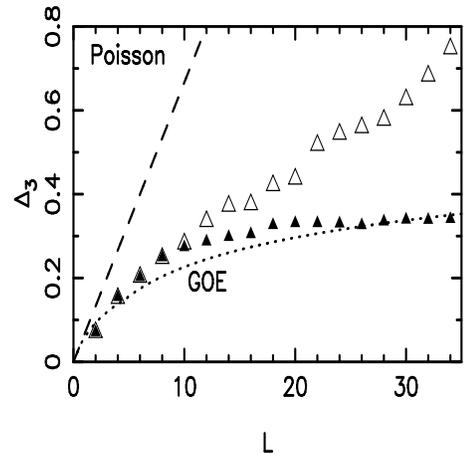}
\end{center}
\caption[]{$\Delta_3$ for the complete $J=10$ level sequence
of a shell-model calculation for $^{52}Ca$ in the $pf$ shell. For
smooth unfolding made by an Edgeworth expansion in the cumulants, the
result (open triangles) lies between Poisson (dashed line) and GOE
(dotted line) limits. Filled triangles correspond to local unfolding
with a $2v=10$ window. }
\label{unfca}
\end{figure}

Let us now consider the atomic nucleus as example of a quantum system
more complex than the quantum billiard. In most nuclei, level
fluctuations are in agreement with GOE predictions at all energies,
showing that the motion is chaotic. However, it has recently been
observed that single closed nuclei are less chaotic than expected
\cite{Otsuka:92,Molina:01}. One of the most regular nuclei at law
energy is $^{52}Ca$. Analysis of the shell-model level spectrum
\cite{Molina:01} shows that the nearest neighbor $P(s)$ distribution
is close to the Poisson limit (the Brody parameter is $\omega=0.25$)
for levels up to 5 MeV above the yrast line. As the excitation energy
is increased, $P(s)$ approaches the spacing distribution of a chaotic
system. However, other statistics indicate that the dynamics still is
not fully chaotic. Often, the mean level density inside a valence
space is very well reproduced by an Edgeworth expansion around a
Gaussian form \cite{Brody:81}.  If we use it to perform a smooth
unfolding of the 2755 $J=10$ levels of $^{52}$Ca, the $\Delta_3(L)$
statistic is close to GOE limit for very small $L$ values, but it
increases linearly instead of logarithmically for larger $L$ values,
as can be seen in Fig. \ref{unfca}. Except for very small $L$ values,
the spectral rigidity is intermediate between that of GOE and Poisson
limits, giving a clear signature of non chaotic motion. This result is
in agreement with the behavior of the wave function localization
lengths \cite{Molina:01}.

When the Edgeworth expansion fails, as it happens sometimes
\cite{Otsuka:92}, the local unfolding or the Gaussian broadening are
the available unfolding methods.  Fig. \ref{unfca} also shows the
results of local unfolding for $v=5$. The calculated $\Delta_3$
follows the line obtained with the smooth unfolding up to $L\simeq
2v$, but then accumulated unfolding errors increase the spectral rigidity
and lead to a $\Delta_3$ saturation for larger $L$. 
This is the same behavior that was
observed in the quantum billiard system described above.  Moreover, as
the $\Delta_3$ values are rather close to the GOE limit when $v=5$ is
used, the conclusion in this case would be that $^{52}$Ca is a chaotic
system.  This example is then very enlightening.  First, it
illustrates that, contrary to common practice, the $\Delta_3(L)$
statistic should be calculated up to high $L$ values, because
otherwise one can miss relevant information on the system
dynamics. Second, it shows the problems that can arise when the mean
level density is not known and one has to rely on local unfolding.

%
%

To avoid any uncertainties on the real mean level density
$\bar{\rho}(E)$ and level fluctuations, we can study GOE and Poisson
level spectra, the paradigmatic cases of chaotic and regular systems,
respectively.  We consider a GOE matrix with dimension $N=10000$, and
compare the spectral fluctuations obtained by three different methods:
Smooth unfolding made with the semicircle law, local unfolding, and
Gaussian broadening unfolding. All these methods yield almost
indistinguishable results for the $P(s)$ distribution, that is in
perfect agreement with the Wigner surmise. The behavior of the short
range correlations is not affected by the method of unfolding.

\begin{figure}[h]
\begin{center}
\leavevmode
\psfig{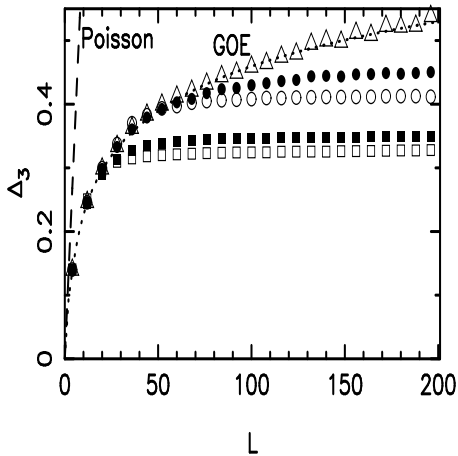}
\end{center}
\caption[]{Comparison of the spectral rigidity for an $N=10000$ GOE
level spectrum, calculated with different unfolding methods. Open
triangles correspond to the smooth unfolding. Filled circles and
squares to the local unfolding with window size $2v=42$ and $2v=18$,
respectively.  Open circles and squares to the Gaussian broadening
unfolding for $\sigma=2$ and $\sigma=1$ MeV, respectively.}

\label{unfgoe}
\end{figure}

However, Fig. \ref{unfgoe} shows a completely different scenario for
$\Delta_3$.  For the smooth unfolding, the spectral rigidity behaves
as predicted by GOE, up to very large $L$ values.  The local unfolding
was performed using two different windows, with $v=9$ and $v=21$. The
calculated $\Delta_3$ coincides now with GOE predictions only up to
$L\simeq 2v$, then it leaves the GOE trend because the spectral rigidity
increases and finally $\Delta_3$ saturates to a constant value. The
Gaussian broadening unfolding was performed for $\sigma=1$ MeV and
$\sigma=2$ Mev. In the central part of the spectrum these values
correspond to windows containing about $10$ and $20$ states,
respectively . Therefore, the effective number of states that affect
the average level density is about the same as in the local unfolding
case. Again, we see the same $\Delta_3$ behavior for $L$ values greater
than the window used in the unfolding.

\begin{figure}[h]
\begin{center}
\leavevmode
\psfig{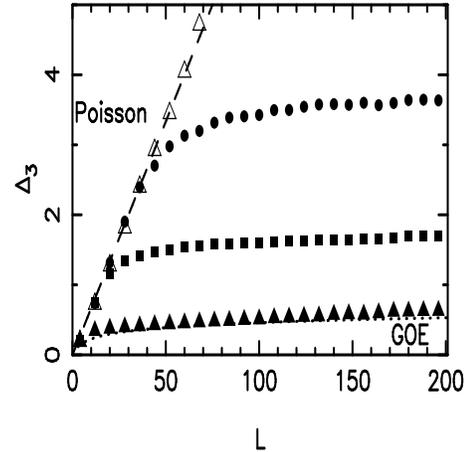}
\end{center}
\caption[]{Comparison of the spectral rigidity for a Poisson sequence of 10000
levels with uniform density, using several unfolding procedures.  Open
triangles correspond to the smooth unfolding. Filled circles, squares
and triangles, to the local unfolding with window size $2v=42$,
$2v=18$ and $2v=4$, respectively.}
\label{unfpois}
\end{figure}

Fig. \ref{unfpois} shows the spectral rigidity for 10000 levels
generated with Poisson statistics and a uniform density
$\bar{\rho}(E)=1$. The smooth unfolding gives $\Delta_3$ values
close to Poisson predictions, but local unfolding with $v=2$, 9 and
21, leads again to the same behavior observed in previous cases 
for $L\agt 2v$. In fact, for the small window with four spacings, 
the $\Delta_3$ of the Poisson spectrum closely follows GOE predictions!

%
%

Looking for deeper insight into the spurious $\Delta_3$ saturation, we
consider the sequence of nearest level spacings as a physical signal,
and apply Fourier analysis techniques to its study.  We have chosen a
system with Poisson statistics and uniform level distribution to
illustrate the idea, because the smooth density is constant.
Therefore, the fundamental assumption of the local unfolding method,
namely that the mean density is approximately linear within a window,
is exactly fulfilled.  From the real nearest-neighbor level spacing
sequence $S$, we obtain: (a) the average spacing sequence $D_L$
calculated with the local constant density of Eq. (\ref{rholoc}) and
$v=21$, (b) the sequence of smoothly unfolded spacings $s$, and (c)
the sequence of locally unfolded spacings $s_L$.  Since for this
spectrum $\bar{\rho}(E)=1$, we have $D=1$ and $s=S$.

\begin{figure}[h]
\begin{center}
\leavevmode
\psfig{file=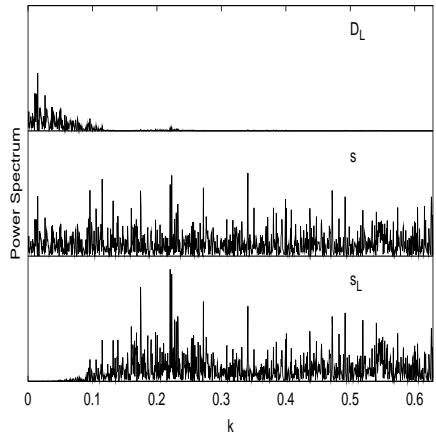,angle=-90,height=6cm,width=6cm}
\end{center}
\caption[]{Comparison of the power spectrum for the sequences $D_L$,
$s$ and $s_L$, for a sequence of 10000 uncorrelated levels with
uniform density.  The locally unfolded spacings $s_L$ are calculated
with a window of size $2v=42$ that corresponds to $k_0=0.15$.}
\label{pow}
\end{figure}

Fig. \ref{pow} displays the power spectrum of these sequences
for frequencies up to $k=0.6$. For $D_L$, it has a maximum near
$k=0$ and decreases smoothly becoming essentially zero at some
threshold frequency $k_0=\pi/v$.  However, this behavior is a spurious
effect, because the real mean spacing $D$ is constant and then its
power spectrum is zero for all the frequencies $k\ne 0$.  Therefore
the local unfolding procedure introduces spurious low frequency
components into the $D_L$ signal.  Comparing the power spectra of
$D_L$ and $s$, it is seen that they are very similar at low
frequencies, except for the damping of the former. The power spectra
of $s$ and $s_L$ are also very similar, except that the low frequency
components are missing in the latter.  These results clarify the
deficiencies of local unfolding. It becomes apparent that the
procedure is filtering out low frequency fluctuations from the
spectrum $s$, and improperly including them in $D_L$. Moreover, by
reducing or eliminating fluctuations of frequency smaller than $k_0$,
the procedure is introducing long range correlations with wave lengths
greater than $2v$.  As this fluctuation reduction is progressive, the spurious
long range correlations become stronger as $L$ increases beyond the
window size $2v$. It is precisely this phenomenon what has
previously been detected by the $\Delta_3$ statistic, that strong long
range correlations leading to a saturation of the $\Delta_3$ are
observed for $L\agt 2v$.

%
%

In summary, we have shown that the correct behavior of $\Delta_3$ is
strongly modified by some commonly used unfolding procedures when the
exact shape of the mean level density is not known. Methods like local
unfolding or Gaussian broadening introduce spurious long range
correlations in the unfolded level spectrum, increasing
the spectral rigidity and leading to a saturation
of $\Delta_3(L)$. In these methods the local average level spacing at
energy $E$ is calculated from the levels inside an energy window
around $E$. The spurious behavior of the $\Delta_3$ statistic  
is observed for $L$ larger than the window
size. In general it gives misleading signatures of quantum chaos,
and for small windows the behavior of $\Delta_3$ may be close
to the GOE limit. Furthermore, the spurious saturation of $\Delta_3$ can
easily be misinterpreted as Berry's saturation.
 
For systems intermediate between regular and chaotic, the traditional
spectral statistics $P(s)$ and $\Delta_3(L)$ for small $L$ values may
be close to the GOE limit, and strong deviations of the spectral
rigidity from GOE predictions only appear for larger $L$ values.  Thus
if the local mean level density is not known from a statistical theory
or a good empirical ansatz, the analysis of energy level fluctuations
will not lead to correct conclusions on the system dynamics. In this
case, it becomes necessary to go beyond level statistics and study
properties of the wave functions, such as localization length,
transition strengths and transition strength sums \cite{Kota:01}.

%
%
We want to acknowledge useful discussions with D. Weinmann. This work
is supported in part by Spanish Government grants for the research
projects BFM2000-0600 and FTN2000-0963-C02.

%
%

\end{multicols}
 
\end{document}